\documentclass[preprint2]{aastex}
\usepackage{graphicx}
\usepackage{epsfig}
\usepackage{lscape}
\usepackage{graphicx}
\usepackage{amsmath}
\usepackage[english]{babel}
\usepackage[titletoc,title]{appendix}
\usepackage{txfonts}
\usepackage{times}
\usepackage{natbib}
\begin{document}

\title{A Stringent Limit on the Warm Dark Matter Particle Masses from  the Abundance of $z=6$ Galaxies in the Hubble Frontier Fields} 
\author{N. Menci$^1$, A. Grazian$^{1}$, M. Castellano$^1$, N.G. Sanchez$^2$ }
\affil{$^1$INAF - Osservatorio Astronomico di Roma, via di Frascati 33, 00040 Monte Porzio Catone, Italy}
\affil{
$^2$Observatoire de Paris, LERMA, CNRS UMR 8112, 61, Observatoire de Paris PSL, Sorbonne Universit\'es, UPMC Univ. Paris 6, 61 Avenue de l'Observatoire, F-75014 Paris, France}
\begin{abstract}
We show that the recently measured UV luminosity functions of ultra-faint lensed galaxies at $z\approx 6$ in the Hubble Frontier Fields provide an unprecedented probe for  the mass $m_X$ of the Warm Dark Matter candidates independent 
of baryonic physics. Comparing the measured abundance of the faintest  galaxies  with the maximum number density of dark matter halos in WDM cosmologies sets a robust limit $m_X\geq 2.9$ keV for the mass of thermal relic WDM particles at a 1-$\sigma$ confidence level, $m_X\geq 2.4$ keV at 2-$\sigma$, and $m_X\geq 2.1$ keV at 3-$\sigma$. 
These constitute the {\it tightest constraints} on WDM particle mass derived to date from galaxy abundance {\it independently of the baryonic physics involved in galaxy formation}. 
 We discuss the impact of our results on the production mechanism of sterile neutrinos. In particular, if sterile neutrinos are responsible for the 3.5 keV line reported in observations of X-ray clusters, our results {\it firmly rule out} the Dodelson-Widrow production mechanism, and yield $m_{sterile}\gtrsim 6.1$ keV for sterile neutrinos produced via the Shi-Fuller mechanism.
 \end{abstract}
\section{Introduction}

In recent years, impressive improvement in the measurement of the faint-end (down to UV magnitudes $M_{UV}\approx -16$) of the galaxy luminosity function (LF) at high redshift $z\gtrsim 6$ has been made possible by the Wide Field Camera 3 (WFC3) on the Hubble Space Telescope (HST), see, e.g., Bouwens et al. (2011, 2015); McLure et al. (2013); Finkelstein et al. (2015). With the Hubble Frontier Fields (HFF) program, 
even fainter galaxies, with intrinsic magnitudes below the HST limits, can be detected thanks to magnification by foreground galaxy clusters.  The HFF program has enabled the detection 
of galaxies with $M_{UV}\approx -15$ at $z\approx 6$  (Atek et al. 2015; Ishigaki et al. 2015) or $M_{UV} \leq -17$ at $z\approx 8$ (Atek et al. 2015a; Ishigaki et al. 2015; Laporte et al. 2015; Castellano et al. 2016a)

Recently, the observations of lensed background galaxies in Abell 2744 and MACS 0416 were used to measure
the luminosity function of galaxies down to ultra-faint magnitudes $M_{UV} =-12.5$ at $z\approx 6$ (Livermore, Finkelstein and Lotz, 2016, LFL16 hereafter). Such measurements have been shown to provide important constraints on the contribution to reionization, and on the star formation and feedback processes of primeval galaxies (Castellano et al. 2016b). However, their potential implication for constraining alternative Dark Matter (DM) models has not been pointed out yet. 

In particular, the observed high
density of galaxies measured at $z\approx 6$ has a deep impact on Warm Dark Matter (WDM hereafter, see Bode, Ostriker \& Turok 2001) models of galaxy formation, based on DM candidates with masses in the keV scale  (de Vega \& Sanchez 2010). In these models, the population of low-mass galaxies is characterized by lower abundances and shallower central density profiles compared to Cold Dark Matter (CDM) due to the dissipation of small-scale density perturbations produced by the free-streaming of the lighter and faster DM particles. Thus, WDM scenarios have been proposed as a solution to some unsolved issues affecting the CDM model on small scales $\lesssim 1$ Mpc, like the steepness of the density profiles in the inner regions of dwarf galaxies (see de Vega, Salucci, Sanchez 2014) and the over-abundance of faint dwarfs around our Galaxy and in our Local Group (see, e.g., Lovell et al. 2012), as well as in the field (Menci et al. 2012; see also Macci\'o et al. 2012; Papastergis et al. 2015). Indeed,  while a refined treatment of baryonic effects entering galaxy formation (in particular feedback from Supernovae) can contribute to solve the problems (see, e.g.,  Governato et al. 2012; Di Cintio 2014)  feedback effects can hardly explain the excess of massive satellite DM halos with virial velocities $V_{vir} \geq 20$ km/s relative to the number of observed bright dwarf galaxies (Boylan- Kolchin et al. 2011), and - most of all - the over-prediction of the abundance of field dwarfs with $V_{vir} \approx $ 40-60 km/s (Klypin et al. 2015). 

The effect of assuming WDM on galaxy formation strongly depends on the mass of the candidate DM particle 
 (see, e.g.,  Polisensky \& Ricotti et al. 2011; Macci\'o et al. 2012; Schneider et al. 2012, Lovell et al. 2012). In fact, 
 the mass of the DM particle determines the suppression of the density power spectrum compared to the CDM case, which drives the formation of cosmic structures.  
 The half-mode mass  $M_{hm}$ - determining the mass scale at which the WDM spectrum is suppressed by 1/2 compared to CDM - is a strong inverse function of the WDM particle mass. Thus, different WDM power spectra are generally labeled in terms of the mass $m_X$ of WDM thermal relic particles, 
 for which a one-to-one correspondence exists between the power spectrum and the particle mass. 
 
Existing astrophysical bounds  on the thermal relic mass $m_X$, have been set by different authors  (e.g., $m_X\geq 2.3$ keV, Polisensky \& Ricotti 2011; $m_X\gtrsim 1.5$ keV,  Lovell et al. 2012;  Horiuch et al. 2014; $m_X\geq 2$, keV Kennedy et al. 2013) by
comparing the predictions from N-body WDM simulation or semi-analytic models with the abundance of observed ultra-faint satellites. Note however that the latter are appreciably sensitive to the assumed completeness corrections (see discussions in Abazjian et al. 2011, Schultz et al. 2014). 
At higher redshifts $z\approx 6$ a limit $m_X\gtrsim 1$ keV has been derived from the UV LFs of faint galaxies ($M_{UV}\approx -16$) by Schultz et al. (2014). Since these approaches are based on the comparison between observed LFs and predicted mass function of DM halos in different WDM models, the delicate issue in these methods is their dependence on the physics of baryons determining the 
mass-to-light 
ratio of faint galaxies. Although to a lesser extent, uncertainties in the baryonic physics also affect (see Garzilli \& Boyarsky 2015 and the discussion in Viel et al. 2013) the tighter constraints achieved so far $m_X\geq 3$ keV, derived by comparing small scale structure in the Lyman-$\alpha$ forest of high- resolution ($z > 4$) quasar spectra with hydrodynamical N-body simulations (Viel et al. 2013). 
An effective way of bypassing the physics of baryons can be found by exploiting the downturn of  
 the halo mass distribution  $\phi (M,z)$ in WDM cosmology at masses close to the half-mode mass scale $M_{hm}$
  (see Schneider et al. 2012, 2013; Angulo et al. 2013; Benson et al. 2013; Paccucci et al. 2013). 
 At any given redshift, the corresponding  maximum number density of halos  $\overline{\phi_{m_X}(z)}\approx \phi(M_{hm}(m_X),z)$ 
  increases with the WDM particle mass (determining the half-mode mass).  
  Thus, measuring galaxy abundances larger than $\overline{\phi_{m_X}(z)}$ at a given redshift sets a lower limit on $m_X$ 
  which is
  {\it completely independent} of the physics of the baryons, since any baryonic effect can only decrease the number of luminous galaxies compared to the number of host DM halos. Such a method is limited by the depth required to measure large galaxy number densities; in fact, 
  such high densities are more easily attained at the faint end of high-redshift LFs.  

 Pacucci et al. (2013) have applied the above procedure to the number density corresponding to two galaxies detected at $z\approx 10$ by the Cluster Lensing And Supernova survey with Hubble (CLASH) obtaining  a lower limit $m_X\geq 0.9$ keV (2$\sigma$). A similar limit has been obtained by Lapi \& Danese (2015) from existing deep UV LFs at $z\approx 7$. A different strategy has been adopted  by Menci et al. (2016), who used abundances obtained at lower $z\approx 2$ from UV LFs of galaxies lensed by the nearby cluster A1689. The ultra-faint magnitudes $M_{UV}\approx -13$ reached by the observed sample allowed to obtain  a lower bound $m_X\geq 1.5$ keV, 
 which is again independent of baryon physics. 
  
Obtaining tighter limits on $m_X$ with the above method requires reaching faint magnitudes $\approx -14$ at  high
redshifts $z\gtrsim 6$.  Thus, the recent measurements of the UV LFs of lensed galaxies 
 down to ultra-faint magnitudes $M_{UV} = -12.5$ at $z\approx 6$ by LFL16 constitute an unprecedented opportunity to derive {\it strong} constraints 
 on the WDM particle mass $m_X$ 
 which are {\it independent of baryonic physics}. 

\section{The Mass Function of Halos in WDM Cosmologies}

The computation of the halo mass function in WDM models is based on the standard procedure 
described and tested against N-body simulations in Schneider et al. (2012, 2013); Benson et al. (2013), Angulo et al. (2013); our computation has been tested against simulations in Menci et al. (2016). Here we provide a brief outline of the main steps.

The key quantity entering the mass function is the variance of the linear power spectrum 
$P(k)$ of DM perturbations (in terms of the wave-number $k=2\pi/r$). Its dependence on the spatial scale $r$ of perturbations is 
\begin{equation} 
{d\,log\,\sigma^2\over d\,log\,r}=-{1\over 2\,\pi^2\,\sigma^2(r)}\,{P(1/r)\over r^3}.
\end{equation}
Here we have used a sharp-$k$ form (a top-hat sphere in Fourier space) for the window function $W(kr)$ relating the variance to the power spectrum $\sigma^2(M)=\int dk\,k^2\,P(k)\,W(kr)/2\,\pi^2$.

Indeed,  in the case of WDM spectra $P(k)$ suppressed at small scales with respect to the scale-invariant CDM behaviour, both theoretical arguments (Benson et al. 2013, Schneider at al. 2013) and comparisons with N-body simulation (see the authors above, and Angulo et al. 2013) impose a sharp-$k$ form (a top-hat sphere in Fourier space) for the window function. In fact, a top-hat filter 
in the real space would result into diverging mass functions for small scales despite the small scale suppression in the power spectrum, due to the 
fact that  longer wavelength modes are getting re-weighted as the mass scale of the filter increases. However, for a sharp-k filter, the normalization $c$ entering the relation between the 
 halo mass $M = 4\pi\,\overline{\rho} (cr)^3/3$ and the  filter scale $r$ must be calibrated through simulations (here $\overline{\rho}$ is the background density of the Universe).  All studies in the literature yield values for $c$  in the range $c = 2.5-2.7$ (see, e.g., Angulo et al. 2013; Benson et al. 2013; Schneider et al. 2013). We shall consider the effect of such an uncertainty on our results. 

In  WDM scenarios the spectrum $P_{WDM}$ is suppressed with respect to the CDM case $P_{CDM}$ below a characteristic scale depending on the mass  $m_X$ of the WDM particles. If WDM is composed 
of
relic thermalized particles, the suppression factor can be parametrized as (Bode, Ostriker \& Turok 2001)
\begin{equation}
{P_{WDM}(k)\over P_{CDM}(k)}=\Big[1+(\alpha\,k)^{2\,\mu}\Big]^{-10/\mu}\,.
\end{equation}
Here the WDM free-streaming scale enters through the quantity 
\begin{equation}
\alpha=0.049 \,
\Bigg[{\Omega_X\over 0.25}\Bigg]^{0.11}\,
\Bigg[{m_X\over {\rm keV}}\Bigg]^{-1.11}\,
\Bigg[{h\over 0.7}\Bigg]^{1.22}\,{h^{-1}\over \rm Mpc},  
\end{equation}
depending on the WDM particle mass. Here $\Omega_X$ is the WDM density parameter, $h$ is the Hubble constant in units of 100 km/s/Mpc, and $\mu=1.12$. A similar expression 
holds for sterile neutrinos provided one substitutes the mass $m_X$ with a mass $m_{sterile}$ adopting proper conversion factors (depending on the assumed neutrino production mechanisms, see, e.g., Destri, de Vega and Sanchez 2013).
Note that the expressions in eqs. (2) and (3) represent fitting formulae for the actual power spectra of thermal relic WDM obtained through the CAMB (Leweis et al. 2000) or the CMBFAST (Seljak \& Zaldarriaga 1996) Boltzmann solver in the case of thermal relics. The difference between the actual Boltzmann solutions and the fitting formulas in eq. (2) and (3) is at percent level (see, e.g., Lovell et al. 2015), with differences in the quantities $\alpha$ and $\mu$ estimated by different authors which are below 5\% (see, e.g., Viel et al. 2005; Hansen et al. 2002). A fitting formula providing an even better agreement with the true numerical solutions has been presented in Destri, De Vega, Sanchez (2013), although it deviates from the form in eqs. (2) and (3) by at most 3\%. 
In our results we shall 
incorporate the above uncertainties, which also enter into the expression for the half-mode mass given below in eq. (5). 
 
The differential halo mass function (per unit $log\,M$) based on the extended Press \& Schechter approach (Bond et al. 1991; Benson et al. 2012; Schneider et al. 2013)  reads
\begin{equation}
 {d\,\phi\over d\,logM}={1\over 6}\,{\overline{\rho}\over M}\,f(\nu)\,{d\,log\,\sigma^2\over d\,log r}\,.
\end{equation}
Here $\nu\equiv \delta_c^2(t)/\sigma^2$ depends on the linearly extrapolated density for collapse in the spherical model $\delta_c=1.686/D(t)$ and $D(t)$ is the growth factor of DM perturbations. We conservatively assume a spherical collapse model for which $f(\nu)=\sqrt{2\nu/\pi}\,exp(-\nu/2)$. 
Assuming an ellipsoidal collapse model would yield lower halo mass function at the low-mass end and - hence - even tighter constraints on the 
WDM particle mass. To adopt a conservative approach, we do not include the effect of residual thermal velocities; in fact, as shown by Benson et al. (2013), this would lead to an increase of the collapse threshold $\delta_c$ at scales below the free-streaming scale resulting on even tighter constraints on $m_X$. 

The mass function in eq. 4 is computed after substituting eq. 1, with a power spectrum $P(k)=P_{WDM}(k)$ 
determined by the WDM particle mass $m_X$ after eqs. 2-3
 (for $P_{CDM}$ we adopt the form  by Bardeen et al. 1986). The resulting mass functions are characterized by a maximum value 
  at masses close to the ‘half-mode’ mass (see Benson et al. 2012; Schneider et al. 2012, 2013;  Angulo et al. 2013; Menci et al. 2016) 
\begin{equation}
M_{hm}={4\pi\over 3}\,\overline{\rho}\,\Bigg[\pi\alpha\Big(2^{\mu/5}-1\Big)^{-{1\over 2\mu}}\Bigg]^{3}\,.
\end{equation}
Correspondingly, the cumulative mass functions saturate to a maximum value  $\overline{\phi_{m_X}(z)}\approx \phi(M_{hm}(m_X),z)$. 
 The dependence of the scale $\alpha$ (eq. 3) on the WDM particle mass $m_X$ yields a half-mode mass ranging from $M_{hm}\approx 10^{10}\,M_{\odot}$ for $m_X=1$ keV   to $M_{hm}\approx 10^{8}\,M_{\odot}$ for $m_X=4$ keV. 

\section{Results}
In fig. 1 we show the cumulative mass function $\phi(>M)$ computed from eq. 4 at $z=6$ for different assumed WDM particle masses, 
 adopting recent Planck cosmological parameters: $\Omega_m=0.32$, $\Omega_{\Lambda}=0.68$, $\Omega_b=0.05$, $h=0.7$, $\sigma_8=0.83$. 
 All the mass functions saturate to a maximum number density $\overline\phi_{m_X}\approx \phi (M_{hm})$. This is compared with the observed number density $\phi_{obs}$ of galaxies with $M_{UV}\leq -12.5$ corresponding to the LFL16 LFs at $z=6$ 
 within 1-$\sigma$,  2-$\sigma$, and 3-$\sigma$ (shaded areas).  
In order to derive the observed cumulative number density $\phi_{obs}$ (and its
confidence levels) corresponding to the differential LFs $\Phi(M_{UV})$ of LFL16, we have 
used the values shown in their Fig. 10 with the corresponding 1-$\sigma$ uncertainties  in each magnitude
bin. We produced Monte Carlo simulations by extracting 
random  values $\Phi_{random}(M_{UV})$ of 
the LF in each magnitude bin according to a Gaussian distribution with 
variance given by the error bar in LFL16.  Thus, for each simulation we produced 
a new realization of the $z=6$ LF. From this, a
cumulative number density $\phi_{random}$ has been derived by summing up
the values of $\Phi_{random}(M_{UV})$ in all the
observed magnitude bins in the range  $-22.5\leq M_{UV}\leq -12.5$.  
We carried out $N_{sim}=10^7$ simulations to compute the probability
distribution function (PDF) of the cumulative number density
$\phi_{random}$. We obtain a median value  $\log\phi_{obs}/Mpc^3=0.54$, 
 while from the relevant percentiles of the PDF we 
 derive lower bounds  0.26, 0.01, and -0.32 at 
1, 2, and 3-$\sigma$ confidence levels, respectively. We have checked that the
median value of the differential LF $\Phi_{random}$ obtained from our simulations  
is consistent (within 3\%) with the best fit value of the LFL16 luminosity function.   
  
 \begin{center}
\vspace{-0.1cm}
\hspace{-0.5cm}
\scalebox{0.4}[0.4]{\rotatebox{-0}{\includegraphics{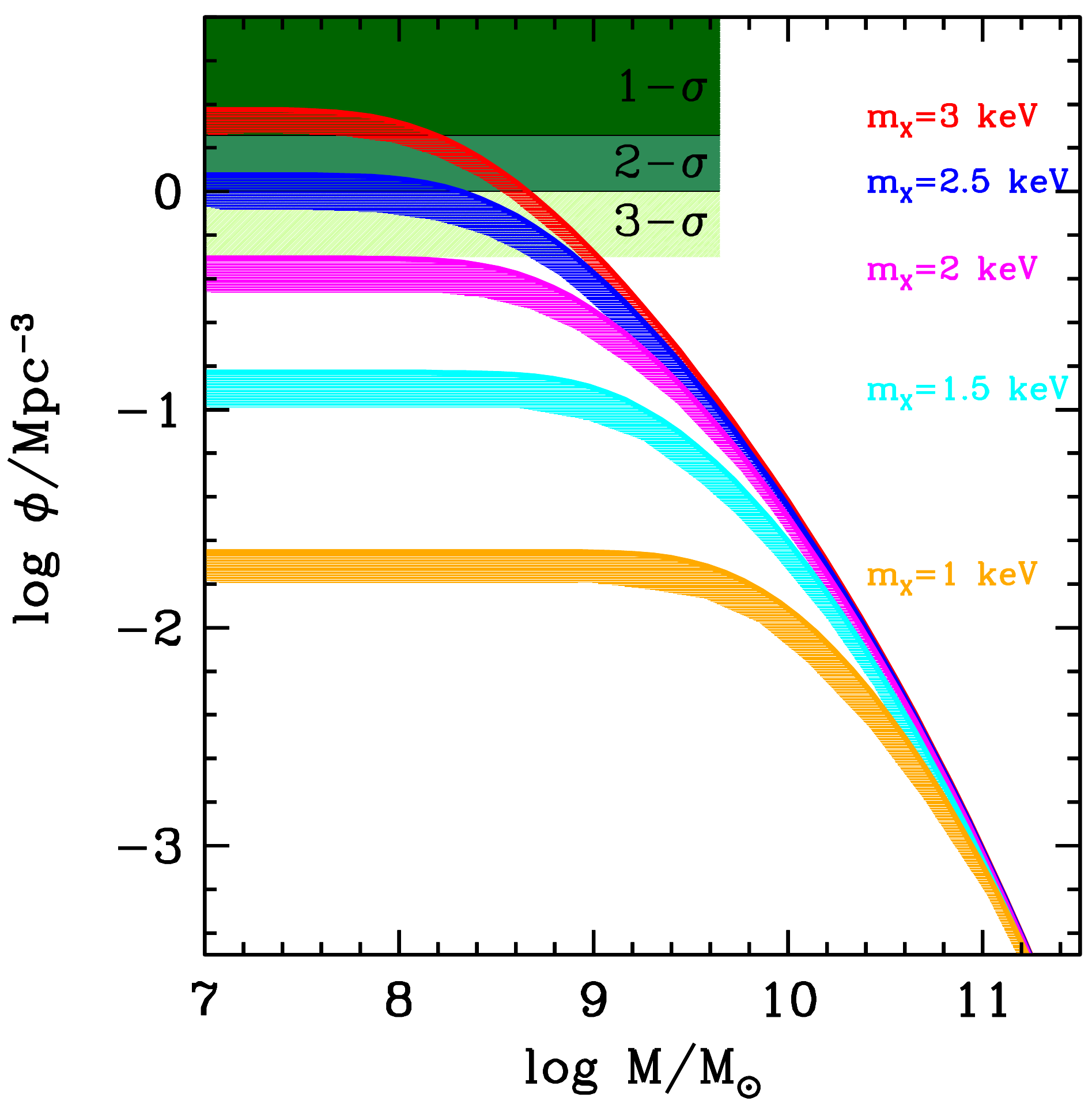}}}
\end{center}
\vspace{-0.3cm }
 {\footnotesize 
Fig. 1. The cumulative mass functions computed at $z=6$ for different values of the WDM particle mass $m_X$ shown by the labels on the right. 
 The thickness of the lines represent the uncertainties in the theoretical predictions related to the window function and to the adopted fitting formula for the WDM power spectrum discussed in sect. 2. 
 The shaded areas correspond to the observed number density of galaxies with $M_{UV}\leq -12.5$ within within 1-$\sigma$, 2-$\sigma$, and  3-$\sigma$  confidence levels. 
\vspace{0.2cm}}

In Figure 2 we compare $\phi_{obs}$ and $\overline\phi_{m_X}$ as a function of $m_X$.
Since luminous galaxies cannot outnumber DM halos, the condition $\phi_{obs}\leq \overline\phi_{m_X}$ yields  $m_X\gtrsim 2.9$ keV at 1-$\sigma$ level,  $m_X\geq 2.4$ keV at 2-$\sigma$ level, and $m_X\geq 2.1$ keV at 3-$\sigma$ level. Our constraints are the 
tightest derived so far from galaxy counts. Although these constraints are less stringent
than the  $2-\sigma$ limit $m_X\geq 3.3$ keV  derived  from the Lyman-$\alpha$ forest  (Viel et al. 2013), our limits are {\it entirely independent} of the modelling of baryons physics which affects the constraints from the Lyman-$\alpha$ absorbers. 

\begin{center}
\vspace{-0.1cm}
\hspace{-0.3cm}
\scalebox{0.4}[0.4]{\rotatebox{-0}{\includegraphics{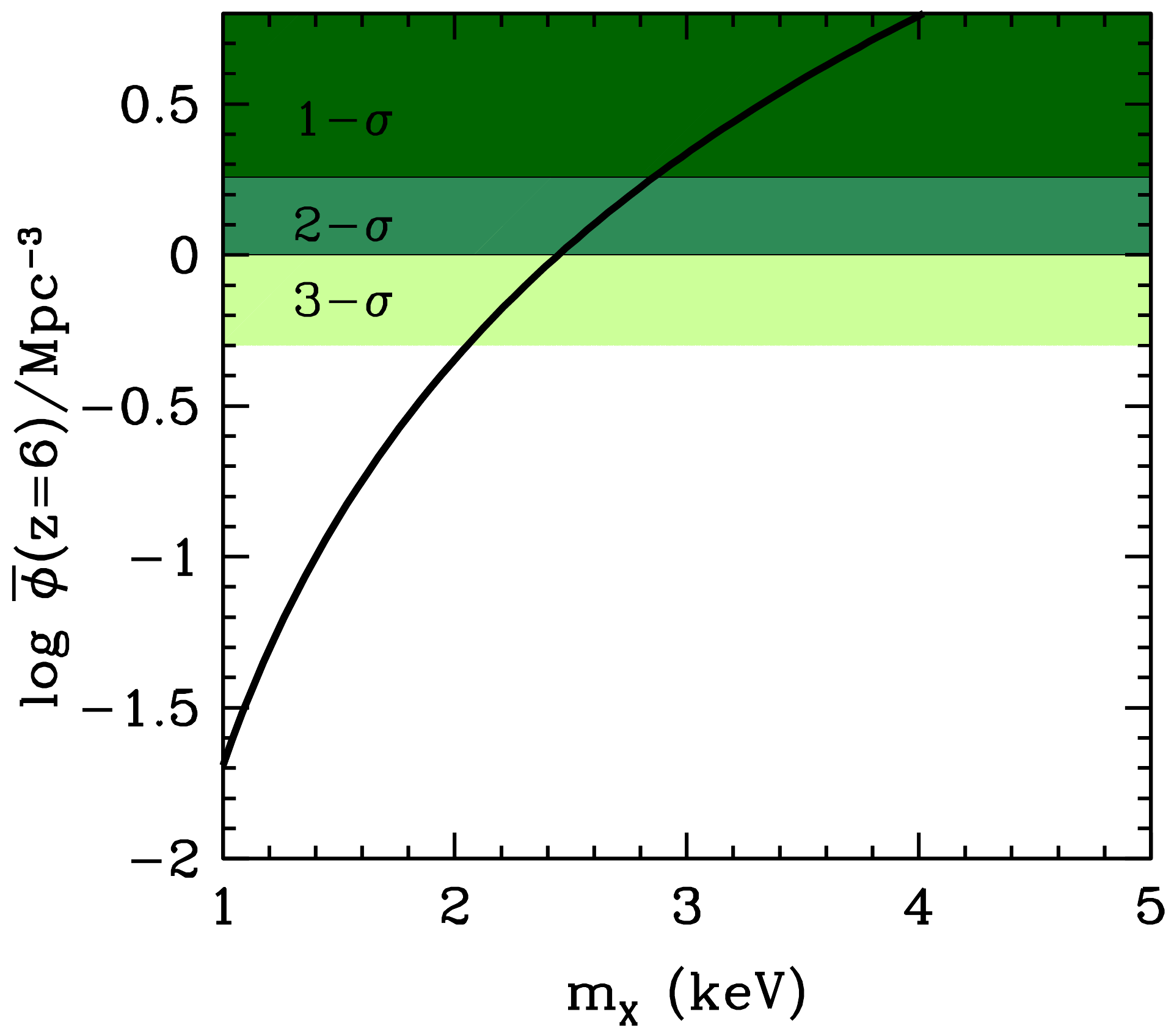}}}
\end{center}
\vspace{-0.3cm }
 {\footnotesize 
 Fig. 2. For different values of the thermal relic mass $m_X$, we show the maximum value (including the theoretical uncertainties) of the predicted number density of  DM halos $\phi$ at $z=6$. 
 The shaded areas represent the observed number density of galaxies with $M_{UV}\leq -12.5$ within 1-$\sigma$, 2-$\sigma$, and  3-$\sigma$  
confidence levels. 
\vspace{0.2cm}}

The method we have applied is similar to that adopted by Pacucci (2013) at $z=10$, by Lapi \& Danese (2015) at $z\approx 7$, and by Menci et al. (2016) at $z\approx 2$. Compared to such works, we derive significantly tighter constraints on $m_X$ due to the unprecedented depth reached by the LF measurements in LFL16. To provide a comparison with previous results and to show how the LFL16 measurements 
made it possible to significantly improve the constraint on 
$m_X$, we show in fig. 3 the thermal relic mass $m_X$ that can be probed by observing a given number density of galaxies $\overline\phi_{m_X}$ (in the y-axis) at different redshifts (x-axis). Such values are compared with the lower bounds set by different measurements at various redshifts. Thus, the contour corresponding to the lower tip of the arrow defines the mass $m_X$ probed by the corresponding observations (at 1-$\sigma$ level). Our $1-\sigma$ lower bound derived from LFL16 is shown by the large circle at $z=6$ and provides the most stringent limit derived so far.

\section{Conclusions}

We show that the recently measured UV luminosity functions (LFs) of ultra-faint lensed galaxies at $z\approx 6$ provide 
unprecedentedly
strong constraints on the mass of WDM candidates $m_X$ 
which is independent 
of baryonic physics. Comparing with the measured abundance of the faintest  galaxies with the maximum number density of dark matter halos in WDM cosmologies sets a robust limit $m_X\geq 2.9$ keV for the mass of thermal relic WDM particles at 1$\sigma$ confidence level,  and $m_X\geq 2.4$ keV at $2-\sigma$ level. 
 
\begin{center}
\vspace{-0.1cm}
\hspace{-0.5cm}
\scalebox{0.45}[0.45]{\rotatebox{-0}{\includegraphics{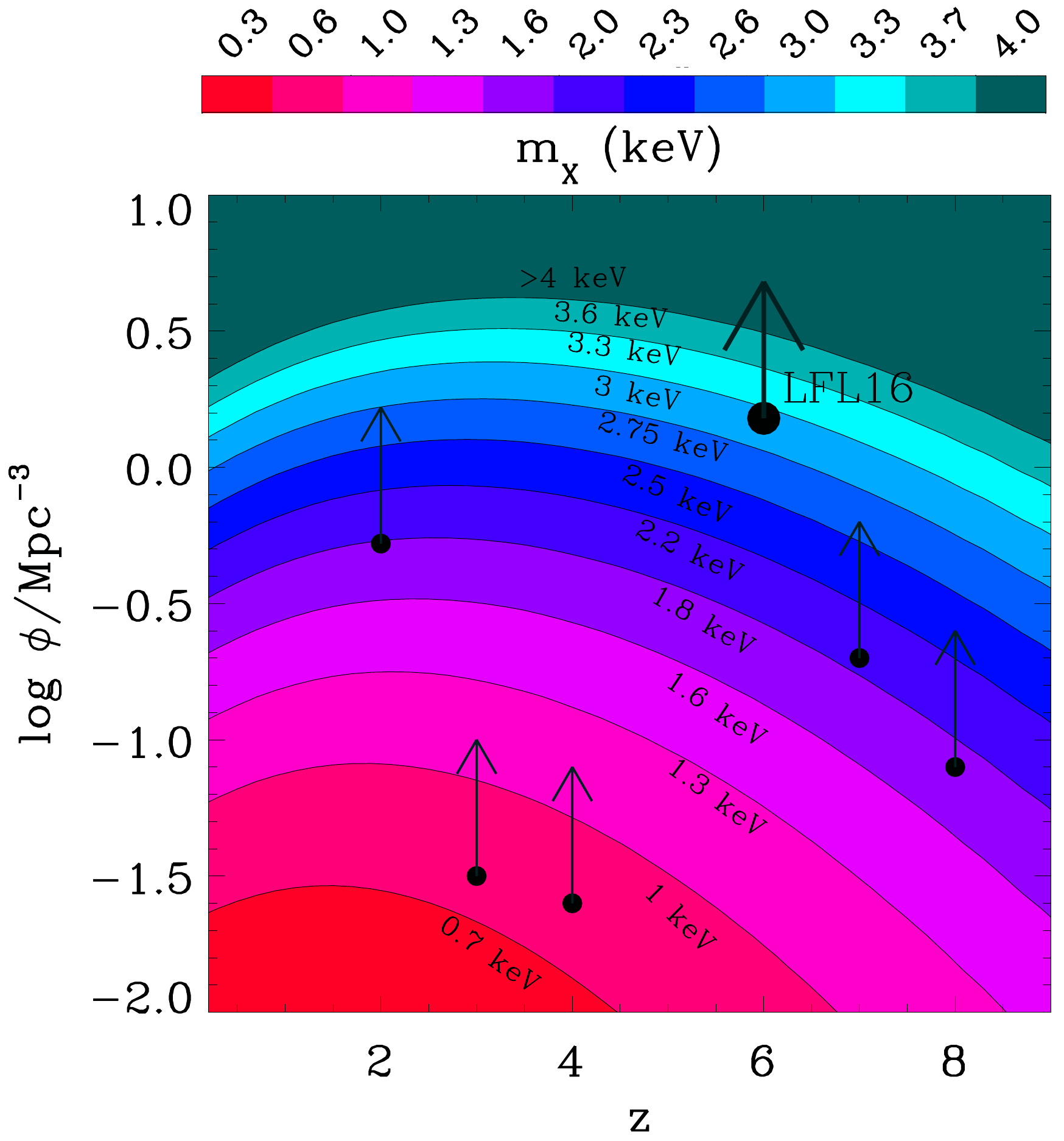}}}
\end{center}
\vspace{-0.3cm }
 {\footnotesize 
Fig. 3. The contours show the maximum number density of DM halos (y-axis) obtained at different redshifts (x-axis) assuming different 
values for the WDM particle mass $m_X$ (contour levels and colors).  Such abundances are compared with the lower limit (at 1-$\sigma$ level) 
set by the different UV galaxy LFs in the literature integrated down to their faintest  magnitude bin at $z=2$ (Alavi et al. et al. 2014), at $z=3-4$ (Parsa et al. 2016); and at $z>6$ by LFL16. The thick dot and error bar corresponds to the UV LFs measured by LFL16 at $z=6$, which provide the 
tightest
bound on $m_X$. 
\vspace{0.2cm}}

 The corresponding lower limit for the sterile neutrino mass depends on the production model. Accurate conversion factors relating the thermal relic mass $m_X$ to the values of $m_{sterile}$ giving the same power spectrum are provided by Destri, de Vega and Sanchez (2013), together with a discussion on the accuracy and comparison of the conversion factors in the literature. They obtain $m_{sterile} \simeq 2.85$ keV $(m_X/{\rm keV})^{4/3}$ for the  Dodelson \& Widrow (1994) mechanism, $m_{sterile}\simeq 2.55\,m_X$ for the Shi \& Fuller (1999) resonant production (for vanishing lepton asymmetry), and  $m_{sterile} \simeq1.9\,m_X$ for the neutrino Minimal Standard Model (Shaposhnikov \& Tkachev 2006). 
  
If sterile neutrinos with mass $m_{sterile}\approx 7$ keV are responsible for the recent unidentified X-ray line at 3.5 keV reported in  observations of X-ray clusters (Bulbul et al. 2014; Boyarsky et al. 2014), our 2-$\sigma$ constraint $m_X\geq 2.4$ keV firmly rules out - independently of astrophysical modelling and of incompleteness corrections - the  Dodelson-Widrow mechanism for the production of sterile neutrinos, already disfavored by previous results (see Horiuchi et al. 2014).    

While our results are robust with respect to astrophysical modelling of baryonic processes involved in galaxy formation, they rely on the 
 observed LFs in LFL16. Such measurements indeed exploit state-of-the-art 
analyses
 of the space
densities of ultra-faint star forming galaxies at $z\ge 6$, thanks to
the very faint limits ($M_{UV}=-12.5$ at $z=6$) reached by deep HST
observations by exploiting the strong lensing magnifications (of a factor
of 50 or more) of the clusters in the HFF campaign. This effect 
makes it possible to
reach luminosities which are more than a factor of 100  deeper
than the ones available in unlensed HST pointings. The strong lensing 
magnifications have been derived by 
adopting the full range of possible lens models produced for the
HFF by seven independent groups who used different
assumptions and methodologies,  to 
check the systematic effects of different lens models on the LFs. 
Notably, LFL16 concluded that the logarithmic slope at faint end varies by
less than 4\% using the different lensing maps available, without any 
turnover down to $M_{UV}=-12.5$ at $z=6$.  
As for the statistics, we note that the large number of
galaxies investigated (167 galaxies at $z=6$) and the availability of
two independent lines of sight also allow LFL16 to reduce the cosmic
variance effect, which usually affects the measurements of the luminosity 
functions at these high redshifts.

However, the results in LFL16 can still be improved both in accuracy
and statistics. First of all through refined methods to assemble the
final source catalog, which is now obtained by a simple positional
cross-matching of 22 independent catalogs derived from different
wavelet images in various bands. Secondly, by reducing the uncertainty
on the photometric redshifts that, as stated in LFL16, dominates over
the variance of different lens models, except in cases where the
magnification is high ($\ge 10$). In this respect, the combination
of different photometric redshift estimates (e.g. Merlin et al. 2016; 
Castellano et al. 2016a) can be a promising way to further reduce the
uncertainties on the LF.

In the future, the current results can be further improved by
extending the present analysis to all the 6 Frontier Field clusters
 (see Lotz et al. 2016). This will make it possible to
reduce the error bars on the LF at
high-z, thanks to the larger number statistics (450 galaxies at $z\sim
6$ are expected) and to the reduced cosmic variance thanks to 6
independent lines of sights.

\begin{acknowledgements}
We warmly thank Anna Nierenberg for substantial help in improving the manuscript.\\ 
\end{acknowledgements}

\end{document}